%% file: main.tex
\documentclass[journal,letterpaper,onecolumn]{IEEEtran}
\IEEEoverridecommandlockouts
\usepackage[utf8]{inputenc}
\usepackage{amsmath,amssymb}
\usepackage{tablefootnote}
\usepackage{multirow}
\usepackage{hyperref}
\input{./preamble}

\title{LLMZip: Lossless Text Compression using Large Language Models}
\author{Chandra Shekhara Kaushik Valmeekam, 
Krishna Narayanan, 
Dileep Kalathil, \\
Jean-Francois Chamberland,
Srinivas Shakkottai \\
Department of Electrical and Computer Engineering \\
Texas A\&M University\\
Email:\{vcskaushik9,krn,dileep.kalathil,chmbrlnd,sshakkot\}@tamu.edu
}

\begin{document}
\maketitle
\begin{abstract}
We provide new estimates of an asymptotic upper bound on the entropy of English using the large language model LLaMA-7B as a predictor for the next token given a window of past tokens. 
This estimate is significantly smaller than currently available estimates in \cite{cover1978convergent}, \cite{lutati2023focus}.
A natural byproduct is an algorithm for lossless compression of English text which combines the prediction from the large language model with a lossless compression scheme. 
Preliminary results from limited experiments suggest that our scheme outperforms state-of-the-art text compression schemes such as BSC, ZPAQ, and paq8h.
\end{abstract}

\section{Introduction}
There are close connections between learning, prediction, and compression. The success of ChatGPT has captured the fascination of general public and brought the connection between learning and prediction to the fore. The main advance brought about by large language models such as LLaMA and GPT-4 is that they excel at predicting the next word (token) in a paragraph based on knowing the past several words (tokens).

The connection between prediction and compression was explored as early as 1951 by Shannon in order to estimate the entropy of the English language \cite{shannon1951prediction}. The idea that a good predictor for the $i$th value in a time series based on the past values can be effectively converted to a good compression algorithm has played a prominent role in information theory.
Many algorithms for speech, image, and video compression exploit this idea either explicitly or implicitly. 
Within the context of lossless compression of English text, the idea of combining a language model with arithmetic coding has emerged
as a very effective paradigm \cite{cleary1984data}.
The performance of such a compression scheme depends substantially on the efficacy of the predictor and
every time there is a major advance in the prediction capability, it behooves us to study its effect on the compression performance. 
Indeed, in 2018, the authors of \cite{goyal2018deepzip} used recurrent neural networks (RNN) as the predictor and reported improved results for certain kinds of sources. 
Their scheme still did not outperform state-of-the-art algorithms such as BSC and ZPAQ for text compression. 

It is therefore natural at this time to study whether we can obtain better compression results and sharper estimates of the entropy of the English language using recent large language models such as LLaMA-7B \cite{touvron2023llama}. This is the main goal of this paper.  
We show that when the LLaMA-7B large language model is used as the predictor,
the asymptotic upper bound on the entropy is 0.709 bits/character
when estimated using a 1MB section of the text8 dataset. 
This is smaller than earlier estimates provided in \cite{cover1978convergent} and \cite[Table 4]{lutati2023focus}.
The estimate of the upper bound increases to 0.85 bits/character for a 100 KB section of the text from \cite{gutenberg_texas}, which is still lower than the estimates in \cite{lutati2023focus}.
When LLaMA-7B is combined with an Arithmetic coder for compression, we obtain a compression ratio of 0.7101 bits/character on a 1MB section of the text8 dataset and a compression ratio of 0.8426 bits/character on a 100KB section of a text from \cite{gutenberg_texas}, which are significantly better than the compression ratio obtained using BSC, ZPAQ and pq8h on the full 100MB of the text8 dataset.

\section{Intuitive explanation of the main idea}
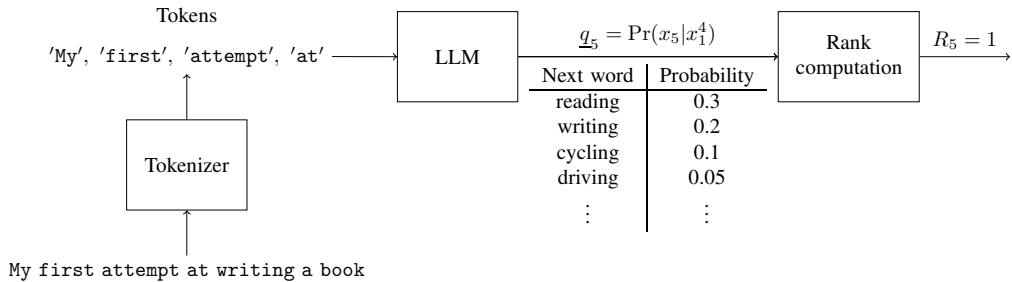
\begin{figure}[htbp]
\begin{center}
\scalebox{0.8}{\input{blockdiagram2}}
\caption{Schematic showing the prediction at epoch 5 for a language model with memory 4.}
\label{fig:nontechexample1}
\end{center}
\end{figure}

We will use the following example to describe the main idea, which is nearly identical to that proposed by Shannon in \cite{shannon1951prediction} for estimating the entropy of English. The main difference is in the use of tokens which represent groups of letters of variable length and in the use of a large language model instead of a human to predict the next token. 
Consider a part of the sentence that reads as
\[
    \tt{My \ first \ attempt \ at \ writing \ a \ book}
\]
Our goal is to convert this sentence into a sequence of bits with the least possible length such that the original sequence can be reconstructed from the sequence of bits.
This sentence can first be split into a sequence of words (tokens)
\[
\tt{'My', \ 'first', \ 'attempt', \ 'at', \ 'writing', \ 'a', \ 'book'}
\] 
A language model with memory $M$ (for example, say $M=4$) predicts the next word in the sentence based on observing the past $M$ words. Specifically, it produces a rank-ordered list of choices for the next word and their probabilities. As shown in Figure~\ref{fig:nontechexample1}, at epoch 5, the model accepts the first 4 words as input and predicts that the next word in the sentence could be words such as $\tt{'reading', 'writing', 'driving', 'cooking'}$ etc.  The main idea is to compute the rank of the actual word in our sentence ($\tt{'writing'}$) in this list and call it $R_5$. 
We will assume that the ranks start at 0 i.e., the most likely word has rank 0, the second most likely word has rank 1, and so on.
In this example, the rank for $\tt{'writing'}$ is $R_5=1$.

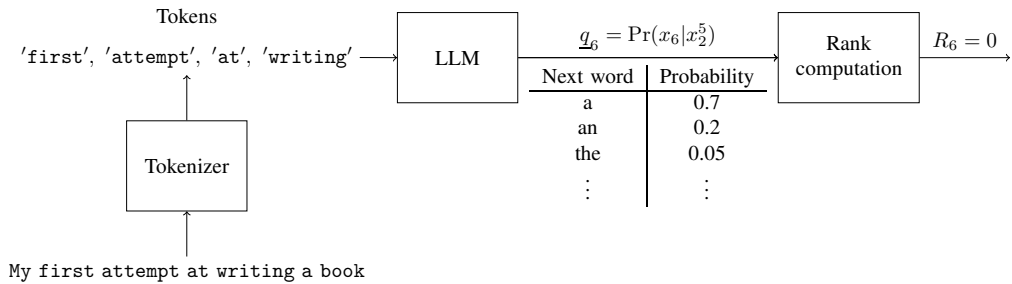
\begin{figure}[htbp]
\begin{center}
\scalebox{0.8}{\input{blockdiagram3}}
\caption{Schematic showing the prediction at epoch 6 for a language model with memory 4.}
\label{fig:nontechexample2}
\end{center}

\end{figure}

 Then, we move forward by one word in the sentence, and at epoch 6, we try to predict the 6th word based on words 2 through 5 as shown in Figure~\ref{fig:nontechexample2}. In this example, given words 2 through 5, the most likely 6th word would indeed be the same word in the sentence that we wish to encode, $\tt{'a'}$, and hence, the rank $R_6$ would be 0.

\begin{figure}[htbp]
\begin{center}
\scalebox{0.8}{\input{ranktobits}}
\caption{Schematic showing the compression of the sequence of ranks to a bit sequence.}
\label{fig:ranktobits}
\end{center}

\end{figure}
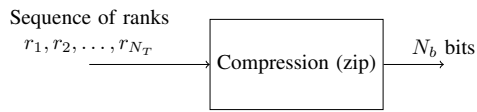

If the language model is good, the word that we wish to encode would often appear at the top of the list and hence, the rank would be 0. Thus, if we look at the sequence of ranks, it is likely to have many $0$s with decreasing probabilities for the rank being $1,2,\ldots$. In this example, it is foreseeable that the ranks will be 
\[
1,0,0,\ldots
\] 
A sequence with many `0's is typically compressible since it has structured patterns. Thus, the key idea is to compress the ranks using a standard lossless compression algorithm such as zip, arithmetic coding, or Huffman coding which converts the ranks to bits. This is shown in Fig.~\ref{fig:ranktobits}.

When we wish to reconstruct the sequence, we first decompress and unzip the bits to get the ranks, use the same language model one epoch at a time to produce a rank ordered list of possibilities for the next word, and pick the word in the list at rank $R_i$ during the $i$th epoch. 
We use that as input for determining the next word and so on. 
Note that this requires that the same LLM is used at both the encoder and the decoder.

The idea of encoding ranks was discussed to build intuition, but better compression can be achieved by directly using the probabilities produced by the LLM along with arithmetic coding as discussed in Section~\ref{sec:LLMAC}.

\section{Compression using LLMs}

Let $\sv$ denote a sentence from the English language composed of $N_c$ letters, where each letter is assumed to be from the alphabet $\mathcal{S}$.
We assume that we have a dictionary $\mathcal{X}=[1,D]$ of $D$ tokens. 
We first parse $\sv$ into a sequence of $N_T$ tokens denoted by $\xv = x_1, x_2, \ldots, x_{i-1}, x_i, x_{i+1}, \ldots x_{N_T}$, where $x_i \in \mathcal{X}$. 
There is a one-to-one mapping between $\sv$ and $\xv$ and hence, compressing $\underline{s}$ is the same as compressing $\xv$.
$x_i$'s can be thought of as realizations of the random variable denoted by the upper case letter $X_i$. 

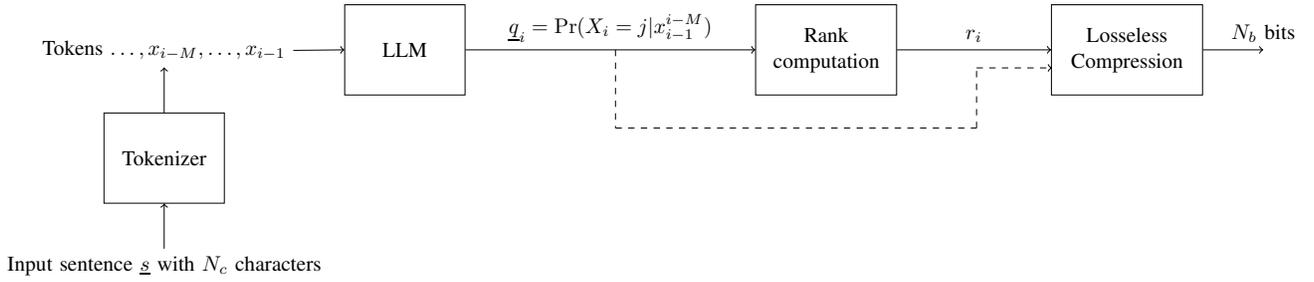
\begin{figure}[htbp]
\begin{center}
\scalebox{0.8}{\input{blockdiagram1}}
\caption{Schematic showing the prediction at epoch $i$.}
\label{fig:techexample}
\end{center}
\end{figure}

A language model with memory $M$ is a predictor that operates as follows. At epoch $i$, it accepts tokens $x_{i-M},x_{i-M+1},\ldots,x_{i-1}$ and produces a probability mass function for the next token in the sequence conditioned on the past $M$ tokens given by $q_i(x_i):= \Pr(X_i = x_i | x_{i-1},x_{i-2}, \ldots, x_{i-M}), \forall x_i \in \mathcal{X}$. The PMF vector $\qv_i:=[q_i(1), q_i(2), \ldots, q_i(D)]^{\mathsf{T}}$ is sorted in descending order and let the sorted PMF vector be denoted by $\tilde{\qv}_i$. Let $\gamma_i:\mathcal{X} \rightarrow \mathcal{X}$ be a permutation on the integers from 1 to $D$ such that 
\[
\tilde{q}_i(\gamma_i(j)) = q_i(j), \forall j \in \mathcal{X}.
\]
That is, $\gamma_i(j)$ is the rank of the token $j$ at epoch $i$.
We define the rank of the input sequence at epoch $i$ as the rank of the token $x_i$ at epoch $i$, $r_i:= \gamma_i(x_i)$.
The sequence $\{r_i\}_{i=1}^{N_T}$ is compressed by a lossless compression algorithm (such as zlib) to produce $N_b$ bits which are the final bit representation of the source.
A schematic of this scheme is shown in Fig.~\ref{fig:techexample}.
In general, the lossless compression algorithm may use the sequence of PMF vectors $\qv_i$'s in addition to the sequence of ranks. 

The main metric of interest is the compression ratio $\rho$ defined as
\[
\rho:=\frac{N_b}{N_c} \hbox{bits/character}.
\]

\subsection{Entropy bounds}

Let $\mathbf{S} \in \mathcal{S}^{\infty}$ be a random process that represents language input.
The $n$th character in the sequence is denoted by $S_n$, whereas the string of characters from the beginning to the $n$th character is expressed as $\mathbf{S}_n$.
The tokenizer parses the input string and maps it to a sequence of tokens $\mathbf{X} = X_1, X_2, \ldots$ using a variable-length mapping.
In this sequence, $X_i$ is the $i$th token.
The number of characters employed to generate $X_i$ depends on the realization of the random process and, as such, we introduce random variable $B_i$ to identify the number of characters contained in the $i$th token.
Motivated by practical considerations, we only admit tokenizers for which $B_i \geq 1$ and $B_i$ is uniformly bounded, with $B_i < \overline{B} < \infty$; these are characteristics of commonly used tokenizers.
An immediate consequence of this framework is that, as the number of tokens grows unbounded $N_T \rightarrow \infty$, the number of characters must also approach infinity $N_c \rightarrow \infty$.
Formally, consider the tokenizer function $T: \mathcal{S}^{\mathbb{N}} \rightarrow \mathcal{X}^{\mathbb{N}}$ operating on infinite symbol sequences; that is, $T (\mathbf{s}) = \mathbf{x}$ where $\mathbf{s}$ is an infinite sequence in $\mathcal{S}^{\infty}$.
For natural number, $i \in \mathbb{N}$, define $m_i : \mathcal{S}^{\mathbb{N}} \rightarrow \mathbb{N}$ to be the (time) index during which the tokenizer working sequentially on an input sequence $\mathbf{s}$ outputs its $i$th token.
Specifically, suppose $\mathbf{s}$ is given, then
\begin{equation} \label{equation:m_i}
m_i (\mathbf{s}) = \min_{n} \left\{ \operatorname{length} \left( T (\mathbf{s}_{n}) \right) \geq i \right\} .
\end{equation}
We note that, by construction, $\lim_{n \rightarrow \infty} \operatorname{length} \left( T (\mathbf{s}_{n}) \right) = \infty$ and, as such, $m_i(\cdot)$ is well-defined.
It may be pertinent to stress that the tokenizer function applied to truncated sequences is not necessarily injective because multiple finite input series can map to the same output.
This phenomenon is a consequence of the fact that, at any point in time, a tokenizer working sequentially may be waiting for an additional symbol before it can unambiguously select the next output token, i.e., there may be instances where $T(\mathbf{s}_{n}) = T(\mathbf{s}_{n+1})$.
However, if we restrict the input series to input indices when a new token is produced, then the restricted mapping becomes injective.
That is, suppose $T(\mathbf{s}) = \mathbf{x}$, then the only (finite) series of input symbols in the restricted set for which $T (\mathbf{y}_n) = \mathbf{x}_i$ is $\mathbf{s}_{m_i (\mathbf{s})}$.
Given a fixed sequence $\mathbf{s}$, we can express the number of characters contained in a token as
\begin{equation*}
b_i = m_i (\mathbf{s}) - m_{i-1} (\mathbf{s})
\end{equation*}
with initial condition $m_{-1} = 0$.
Consequently, the number of characters embedded in the first $N_T$ tokens for a random input becomes $N_c = \sum_{i=1}^{N_T} B_i$.

Having established these properties, we turn to the relation between $H(\mathbf{S})$ and $H(\mathbf{X})$.
We make the assumption that 
$\{S_k\}_{k=1}^{\infty}$,
$\{B_i\}_{i=1}^{\infty}$, and $\{X_i\}_{i=1}^{\infty}$ are stationary and ergodic processes. 
We know from the Shannon-McMillan-Breiman Theorem
\cite{cover1999information} that
\begin{equation} \label{equation:SMB-S}
- \frac{1}{n} \log_2 p_{\mathbf{S}_n}(S_1, \ldots, S_n)
= - \frac{1}{n} \log_2 p_{\mathbf{S}_n}(\mathbf{S}_n)
\rightarrow H(\mathbf{S}) \quad {\text{almost surely}} .
\end{equation}
Let $\Omega_{\mathbf{S}}$ be the collection of $\omega \in \Omega$ for which this limit holds.
In an analogous manner, the Shannon-McMillan-Breiman theorem implies
\begin{equation} \label{equation:SMB-X}
- \frac{1}{i} \log_2 p_{\mathbf{X}_i}(X_1, \ldots, X_i)
= - \frac{1}{i} \log_2 p_{\mathbf{X}_i}(\mathbf{X}_i)
\rightarrow H(\mathbf{X}) \quad {\text{almost surely}} .
\end{equation}
Define $\Omega_{\mathbf{X}}$ as the collection of $\omega \in \Omega$ for which this limit holds.
Finally, by construction, we have
\begin{equation} \label{equation:LLN-B}
\lim_{i \rightarrow \infty} \frac{m_i (\mathbf{S})}{i} = \mathbb{E} \left[ B \right] \quad {\text{almost surely}} .
\end{equation}
Set $\Omega_B$ to be the set of $\omega \in \Omega$ for which this limit holds.
For any $\omega \in \Omega_{\mathbf{S}} \cap \Omega_{\mathbf{X}} \cap \Omega_B$, we deduce that
\begin{equation*}
\begin{split}
H(\mathbf{S})
&= \lim_{k \rightarrow \infty} - \frac{1}{k} \log_2 p_{\mathbf{S}_k}(\mathbf{S}_k (\omega)) \\
&= \lim_{i \rightarrow \infty} - \frac{1}{l_i} \log_2 p_{\mathbf{S}_{l_i}}(\mathbf{S}_{l_i} (\omega)) \\
&= \lim_{i \rightarrow \infty} - \frac{1}{l_i} \log_2 \Pr \left( \mathbf{X}_i = T (\mathbf{S}_{l_i} (\omega)) \right) \\
&= - \frac{1}{\mathbb{E} [B]} \lim_{i \rightarrow \infty} \frac{1}{i} \log_2 \Pr \left( \mathbf{X}_i = \mathbf{x}_i \right)
= \frac{H(\mathbf{X})}{\mathbb{E} [B]} .
\end{split}
\end{equation*}
The first equality follows from \eqref{equation:SMB-S}.
The second equality is a consequence of the fact that $\{ l_i = m_{i} (\mathbf{S}(\omega)) | i \in \mathbb{N} \}$ is an infinite subset of the natural numbers.
Since a subsequence of a convergent sequence must converge to the same limit, we immediately gather that this alternate form approaches $H(\mathbf{S})$.
The third equality is a consequence of the equivalence between the following two events,
\begin{equation*}
\{ \omega \in \Omega | \mathbf{X}_i (\omega) = \mathbf{x}_{i} \}
= \{ \omega \in \Omega | T (\mathbf{S}_{m_{i} (\mathbf{S}(\omega))} ) = \mathbf{x}_{i} \} .
\end{equation*}
This is characteristic of the tokenization process, and it is a consequence of the correspondence described above.
The last step holds because we are considering an $\omega \in \Omega_B$.
The sets $\Omega_{\mathbf{S}}$, $\Omega_{\mathbf{X}}$, and $\Omega_B$ each have probability one; this implies that their intersection also has probability one,
Thus, we must conclude that
\begin{equation*}
H(\mathbf{S}) = \frac{H(\mathbf{X})}{\mathbb{E} [B]} \quad {\text{almost surely}} .
\end{equation*}
As a corollary to this result, any upper bound on $H(\mathbf{X})$ produces an upper bound on $H(\mathbf{S})$.
This is the property we wish to exploit.

Then,
from the results of \cite{cover1978convergent}, we can see that
\begin{equation}
    \Pr \Bigg \{ H(\Xv)  \leq  \lim_{N_T \rightarrow \infty} -\frac{1}{N_T} \sum_{i=1}^{N_T} \log_2 q_i(X_i)
   \Bigg \} = 1,
\end{equation}
where $q_i(\cdot)$ is the output PMF from the language model.
Therefore, an asymptotic upper bound on the entropy rate $H(\Sv)$ is given by
\begin{equation}
\label{eqn:asymptoticupperbound} 
H(\Sv)\leq \frac{ \lim_{N_T \rightarrow \infty} 
    -\frac{1}{N_T}\sum_{i=1}^{N_T} \log_2 q_i(X_i)}{\mathbb{E}[B]}.
\end{equation}


We refer to the expression in the right hand side of \eqref{eqn:asymptoticupperbound} as the asymptotic upper bound on $H(\Sv)$ and denote it by $H_{ub}$.
The numerator in \eqref{eqn:asymptoticupperbound} represents the average number of bits required to represent the tokens $\mathbf{X}_{N_T}$ and the denominator in \eqref{eqn:asymptoticupperbound} is the average number of charcaters per token.
Hence, the unit for $H(\Sv)$ is bits/character.
In \cite{cover1978convergent}, Cover and King provide 1.3 bits/character as an estimate of the asymptotic upper bound on $H(\Sv)$. 
They also provide an  extensive list of references and discussion of the literature on estimating the entropy of English prior to 1976.
Very recently, in \cite[Table 4]{lutati2023focus}, the performance of several language models have evaluated on the text8 dataset using a metric called bits per character (bpc).
We believe bpc is the same as the asymptotic upper bound in this paper.


\subsection{Encoding schemes}
We consider three schemes for the lossless compression block in Fig.~\ref{fig:ranktobits}.

\subsubsection{Compressing the ranks using zlib}
The first scheme uses the zlib compression algorithm to encode the sequence of ranks. We refer to this scheme as LLaMA+zlib and denote the compression ratio of this scheme by $\rho_{\text{LLaMA+zlib}}$.

\subsubsection{Token-by-Token Compression}
The second scheme uses a token-by-token lossless compression scheme which uses a time-varying codebook to encode the token $x_i$ at epoch $i$ by using a prefix-free code assuming $q_i$ to be the true distribution of the tokens. 
A natural choice for a prefix-free code is a Huffman code. 
Instead, for simplicity, we
use a prefix-free code where the codeword for the token $x_i$ is of length $l_i = \lceil \log_2 \frac{1}{q_i(x_i)} \rceil$. 
A prefix-free code with this length for $x_i$ is guaranteed to exist since this choice of lengths satisfies the Kraft inequality \cite{cover1999information}.
The compression ratio for this scheme, denoted by 
$\rho_{\text{LLaMA+TbyT}}$, is given by
\[
\rho_{\text{LLaMA+TbyT}} = \frac {\displaystyle{\sum_{i=1}^{N_T} \left \lceil \log_2 \frac{1}{q_i(x_i)} \right \rceil}}
{\sum_{i=1}^{N_T} b_i}.
\]

\subsubsection{Arithmetic Coding}
\label{sec:LLMAC}
The above two schemes are intuitive but their performance can be improved. 
A very effective way to combine the output of the LLM with a lossless compression scheme is by using arithmetic coding \cite{cleary1984data,bell1989modeling}. 
Arithmetic coding is well suited to accept time-varying probabilities and we use $q_i(x_i)$ as the probability of token $x_i$ at time in the arithmetic coding scheme.
We refer to the compression ratio of this scheme as
$\rho_{\text{LLM+AC}}$.
It is known that arithmetic coding is nearly optimal as a compression scheme \cite[Page 115]{mackay2003information}. 
Hence, the compression ratio for this scheme is expected to be 
\begin{equation}
    \label{eqn:rhoLLaMA+AC}
    \rho_{\text{LLM+AC}} \approx \frac {\displaystyle{\sum_{i=1}^{N_T} \log_2 \frac{1}{q_i(x_i)}}}
{\sum_{i=1}^{N_T} b_i}.
\end{equation}

Clearly, $\rho_{\text{LLaMA+zlib}}$, $\rho_{\text{LLaMA+TbyT}}$, and 
$\rho_{\text{LLM+AC}}$
also provide upper bounds on $H(\Sv)$.
$H_{ub}, \rho_{\text{LLaMA+zlib}}$, $\rho_{\text{LLaMA+TbyT}}$,  and $\rho_{\text{LLM+AC}}$
are estimated using a finite number of tokens and the statistical properties of such an estimate should be kept in mind when interpreting the results, especially since the tokens are from a very large alphabet and language model has large memory.


\section{Results}
We used LLaMA-7B \cite{touvron2023llama} as the large language model and SentencePiece tokenizer \cite{kudo2018sentence}. The tokenizer produces a dictionary of size 32000.
Since the language model is trained on this tokenizer, it is imperative that this tokenizer be used in conjunction with the LLM.
It should be noted that the tokenizer and the model are trained on a large corpus of text which includes uppercase letters, special characters etc.
This is in contrast to many studies on estimating the entropy of English, where the input alphabet is restricted to lowercase letters such as in \cite{shannon1951prediction, cover1978convergent, goyal2018deepzip}.
This makes it difficult to perform an entirely fair comparison between these models. 
By using a pretrained LLM on an input consisting only of lowercase letters, we may be unfair to the LLM. 

Nevertheless, we used the text8 dataset available from 
\url{http://mattmahoney.net/dc/text8.zip} to benchmark the performance of LLaMA-7B with compression against other state of the art results for text compression. 
In \cite{goyal2018deepzip}, it is mentioned that the ZPAQ algorithm obtains the best compression ratio for the text8 dataset with a compression ratio of 1.4 bits/character. 
In \cite{text8results}, the paq8h algorithm is shown to provide a compression ratio of 1.2 bits/character.
To the best of our knowledge, this appears to be best performance reported.
Therefore, we used these two algorithms as baselines.
We did not independently run the ZPAQ or paq8h algorithms and we are quoting results from the existing literature.

The performance of LLaMA-7B is shown in Table~\ref{table:LLaMAresults512} for 10 different batches each with 100,000 tokens. The average performance over these 1M tokens is also shown in the last row in the Table. 
It can be seen that using LLaMA-7B with Arithmetic Coding compression results in a compression ratio of 0.7101 bits/character. 
This is substantially better than the state-of-the-art results mentioned in \cite{goyal2018deepzip} or \cite{text8results} and is very close to our computed upper bound.
The performance with the LLaMA+zlib algorithm and LLaMA+TbyT compression are also better than that of the known state-of-the-art results.
Table~\ref{table:LLaMAresults512} also shows the upper bound in \eqref{eqn:asymptoticupperbound}. 
It should be noted that the upper bound on the entropy is lower than that computed by Shannon in \cite{shannon1951prediction}, Cover and King in \cite{cover1978convergent} and more recent estimates based on neural networks in \cite{lutati2023focus}.

The dependence of the compression performance on the memory of the LLM ($M$) is shown in Table \ref{table:LLaMAresults_memory}. 
As expected, the compression performance improves with increasing $M$. We also observed that the inference time scaled approximately linearly with the input memory length, i.e., batches with a memory of 511 tokens ran about 16 times slower than batches with a memory of 31 tokens.

It is well known that the estimate of compression ratio can show substantial variance depending on the input text and hence, the results should be interpreted with caution. The empirical mean and standard deviation of the entropy bounds and compression ratios computed using 10 batches of 100,000 tokens are shown in Table \ref{table:LLaMAresults_memory_var}. 
We were also not able to run LLaMA-7B on the entire 100MB of the text8 dataset. So, the comparison of LLaMA-7B with that of the state-of-the-art corresponds to estimates obtained from different input sizes.

It appears that the LLaMA-7B model was trained on a corpus that included articles from Wikipedia. 
Since the text8 dataset is derived from Wikipedia, it is likely that our results for the text8 dataset are optimistic.


Therefore, we also tested the performance of LLaMA-7B on a recently released (May 25, 2023) book \cite{gutenberg_texas} under Project Gutenberg. We extracted text that corresponds to 100,000 tokens.  We applied the same text pre-processing as used in the text8 dataset to clean the text from the book. The resulting text data contained only lowercase letters and space as in the text8 dataset. 
Table \ref{table:LLaMAresults_memory_gutenberg} shows the compression performance of the LLM on the book.
It can be seen that the compression ratios and the entropy upper bound are slightly higher compared to the performance on the text8 dataset; nevertheless, the asymptotic upper bound on the entropy is lower than that of currently known models given in \cite[Table 4]{lutati2023focus}). 
Similarly, the compression ratio of LLaMA-7B-based compressors are better than those of known state-of-the-art results for the text8 dataset.
The compression ratio for LLaMA with arithmetic coding is only 0.8426 bits/character and is very close to the estimated upper bound on $H(\Sv)$.

To provide some insight into the comparative performance of LLaMA based compressors vis-a-vis standard text compressors, we also ran the zlib algorithm directly on the input text. 
The resulting compression ratio was 2.8 bits/character (shown in the last column).
It is clear that the performance of LLaMA based compressors is substantially better than this. 
The zlib algorithm may not be optimized for compressing small text samples and hence, the compression ratio for the zlib algorithm and the LLaMA+zlib will likely improve on longer texts.

\section{Acknowledgement}
We would like to thank Andreas Kirsch for an email discussion about arithmetic coding that motivated us to add our results on arithmetic coding in a timely manner.

\begin{table}[h]
    \centering
    \caption{Results for 1MB of text from text8 dataset}
    \label{table:LLaMAresults512}
    \def\arraystretch{1.5}%
    \begin{tabular}{|c|c|c|c|c|c|c|c|c|}
    \hline
    Batch & \multirow{2}{*}{\centering $N_c$} & \multirow{2}{*}{\centering $N_T$} & $H_{\text{ub}}$ & $\rho_{\text{LLaMA+zlib}}$ & $\rho_{\text{LLaMA+TbyT}}$ & $\rho_{\text{LLaMA+AC}}$  & ZPAQ & pq8h \\ 
    No. &  &  & (bpc) & file size (bits) & (bpc) & (bpc) & (bpc) & (bpc) \\
    \hline
    1 & $466,650$ & $100,000$ & $0.6882$ & $1.0513$ & $0.8215$ & $0.689$ & & \\
    \hline
    2 & $461,477$ & $100,000$ & $0.6893$ & $1.0558$ & $0.8242$ & $0.6901$ & &\\
    \hline
    3 &$454,599$ & $100,000$ & $0.699$ & $1.0681$ & $0.8357$ & $0.6999$ & &\\
    \hline
    4 &$462,755$ & $100,000$ & $0.6748$ & $1.0346$ & $0.8093$ & $0.6757$ & &\\
    \hline
    5 &$453,847$ & $100,000$ & $0.7481$ & $1.1265$ & $0.8831$ & $0.749$ & &\\
    \hline
    6 &$458,252$ & $100,000$ & $0.7218$ & $1.0957$ & $0.8567$ & $0.7227$ & &\\
    \hline
    7 &$451,036$ & $100,000$ & $0.6959$ & $1.0729$ & $0.8353$ & $0.6968$ & &\\
    \hline
    8 &$447,953$ & $100,000$ & $0.7092$ & $1.0896$ & $0.8489$ & $0.7101$ & &\\
    \hline
    9 &$462,665$ & $100,000$ & $0.7394$ & $1.1126$ & $0.8713$ & $0.7402$ & &\\
    \hline
    10 &$449,621$ & $100,000$ & $0.7269$ & $1.1046$ & $0.8643$ & $0.7277$ & & \\
    \hline
    Total & $9,137,710$ & $2,000,000$ & $0.7093$ & $1.0812$ & $0.845$ & $0.7101$ & 1.4\tablefootnote{This result is taken from \cite{goyal2018deepzip} and it corresponds to the full 100MB dataset text8} & 1.2\tablefootnote{This result is taken from \cite{text8results} and it corresponds to the full 100MB dataset text8} \\
    \hline
    \end{tabular}
\end{table}

\begin{table}[h]
    \centering
    \caption{Compression performance of the LLM on the text8 dataset, as a function of its memory ($M$) }
    \label{table:LLaMAresults_memory}
    \def\arraystretch{1.5}%
    \begin{tabular}{|c|c|c|c|c|c|c|c|c|}
    \hline
    \multirow{2}{*}{\centering $M$} & \multirow{2}{*}{\centering $N_c$} & \multirow{2}{*}{\centering $N_t$} & $H_{\text{ub}}$ &  $\rho_{\text{LLaMA+zlib}}$ & $\rho_{\text{LLaMA+TbyT}}$ & $\rho_{\text{LLaMA+AC}}$ \\ 
     &  &  & (bpc) & file size (bits) & (bpc) & (bpc)\\
    \hline
     $31$ & $4,568,855$ & $1,000,000$ & $0.9139$ & $1.3159$ & $1.0425$ & $0.9145$ \\
    \hline
    $127$ & $4,568,855$ & $1,000,000$ & $0.7511$ & $1.1303$ & $0.8847$ & $0.752$ \\
    \hline
    $255$ & $4,568,855$ & $1,000,000$ & $0.7242$ & $1.0985$ & $0.859$ & $0.725$ \\
    \hline
    $511$ & $4,568,855$ & $1,000,000$ & $0.7093$ & $1.0812$ & $0.845$ & $0.7101$ \\
    \hline
    \end{tabular}
\end{table}

\begin{table}[h]
    \centering
    \caption{Mean and standard deviation of the entropy bounds measured over 10 batches of 100,000 tokens}
    \label{table:LLaMAresults_memory_var}
    \def\arraystretch{1.5}%
    \begin{tabular}{|c|c|c|c|c|}
    \hline
    \multirow{2}{*}{\centering $M$} & $ H_{\text{ub}}$ &   $\rho_{\text{LLaMA+zlib}}$ & $\rho_{\text{LLaMA+TbyT}}$ & $\rho_{\text{LLaMA+AC}}$ \\ 
     &  (bpc) & (bpc) & (bpc) &(bpc) \\
    \hline
     $31$ & $0.9139 \pm 0.0263$ & $1.3159 \pm 0.0329$ & $1.0425 \pm 0.0262$ & $0.9145 \pm 0.0263$ \\
\hline
$127$ & $0.7511 \pm 0.0233$ & $1.1303 \pm 0.0292$ & $0.8847 \pm 0.0231$ & $0.752 \pm 0.0233$ \\
\hline
$255$ & $0.7242 \pm 0.0234$ & $1.0985 \pm 0.0289$ & $0.859 \pm 0.0232$ & $0.725 \pm 0.0234$ \\
\hline
$511$ & $0.7093 \pm 0.0228$ & $1.0812 \pm 0.028$ & $0.845 \pm 0.0226$ & $0.7101 \pm 0.0228$ \\
\hline
    \end{tabular}
\end{table}

\begin{table}[h]
    \centering
    \caption{Compression performance of the LLM on a recently published book in Project Gutenberg \cite{gutenberg_texas}, as a function of its memory ($M$) }
    \label{table:LLaMAresults_memory_gutenberg}
    \def\arraystretch{1.5}%
    \begin{tabular}{|c|c|c|c|c|c|c|c|c|c|}
    \hline
    \multirow{2}{*}{\centering $M$} & \multirow{2}{*}{\centering $N_c$} & \multirow{2}{*}{\centering $N_t$} & $H_{\text{ub}}$ & $\rho_{\text{LLaMA+zlib}}$ & $\rho_{\text{LLaMA+TbyT}}$ & $\rho_{\text{LLaMA+AC}}$ & Standalone Zlib \\ 
     &  &  & (bpc) & (bpc) & (bpc) & (bpc) & (bpc)\\
    \hline
     $31$ & $508,463$ & $115,000$ & $1.0919$ & $1.5316$ & $1.2152$ & $1.0924$ & $2.80$ \\
    \hline
    $127$ & $508,463$ & $115,000$ & $0.8973$ & $1.3128$ & $1.0235$ & $0.8982$ & $2.80$\\
    \hline
    $255$ & $508,463$ & $115,000$ & $0.8618$ & $1.2684$ & $0.9899$ & $0.8627$  & $2.80$ \\
    \hline
    $511$ & $508,463$ & $115,000$ & $0.8417$ & $1.2465$ & $0.9711$ & $0.8426$ & $2.80$\\
    \hline
    \end{tabular}
\end{table}

\clearpage

\clearpage
\bibliographystyle{IEEEbib}
\bibliography{references}

\end{document}

%% file: preamble.tex
\usepackage{amsmath,amssymb,amsthm}
\usepackage{enumerate}
\usepackage{stfloats}
\usepackage{comment}
\usepackage{bbm}
\usepackage{siunitx}
\usepackage{mathtools}
\usepackage{accents,latexsym,cancel}
\usepackage{cite}

\usepackage[dvipsnames]{xcolor}
\definecolor{myPink}{RGB}{255,105,183}

\usepackage[T1]{fontenc}
\usepackage{graphics} 
\usepackage{epsfig} 
\usepackage[mathscr]{euscript}
\usepackage{algorithm}
\usepackage[noend]{algpseudocode}
\usepackage{bbm}
\makeatletter
\def\BState{\State\hskip-\ALG@thistlm}
\makeatother

\usepackage{tikz}
\usetikzlibrary{arrows,shapes,chains,matrix,positioning,scopes,patterns,calc,
decorations.markings,
decorations.pathmorphing,
}

\usepackage{pgfplots}
\pgfplotsset{compat=1.3}
\usepgflibrary{shapes}

\renewcommand{\epsilon}{\varepsilon}

\newcommand{\RNum}[1]{\uppercase\expandafter{\romannumeral #1\relax}}

\newcommand{\qv}{\ensuremath{\mathbf{q}}}

\newcommand{\sv}{\ensuremath{\mathbf{s}}}
\newcommand{\Sv}{\ensuremath{\mathbf{S}}}

\newcommand{\xv}{\ensuremath{\mathbf{x}}}
\newcommand{\Xv}{\ensuremath{\mathbf{X}}}

\def\Pr{\mathrm{Pr}}

\DeclareMathAlphabet{\mcl}{OMS}{cmsy}{m}{n}

\newlength\tikzwidth
\newlength\tikzheight

\definecolor{mycolor1}{rgb}{0.63529,0.07843,0.18431}%
\definecolor{mycolor2}{rgb}{0.00000,0.44706,0.74118}%
\definecolor{mycolor3}{rgb}{0.00000,0.49804,0.00000}%
\definecolor{mycolor4}{rgb}{0.87059,0.49020,0.00000}%
\definecolor{mycolor5}{rgb}{0.00000,0.44700,0.74100}%
\definecolor{mycolor6}{rgb}{0.74902,0.00000,0.74902}%

{\begin{list}{$\bullet$}
   {\setlength{\itemsep}{0in}
    \setlength{\labelsep}{0.05in}
    \setlength{\labelwidth}{0.2in}
    \setlength{\leftmargin}{0.25in}
    \setlength{\rightmargin}{0in}
    \setlength{\topsep}{0in}
   }}
{\end{list}}

\usepackage[normalem]{ulem}

%% file: blockdiagram2.tex
\begin{tikzpicture}
[node distance=2cm, block/.style={draw, minimum width=2cm, minimum height=1.5cm}]

\node[block] (input) {Tokenizer};
\draw[->] (input.90) -- ++(0,0.75) node[above] (tokenstext) 
{$\tt{'My',\ 'first',\ 'attempt',\ 'at'}$};
\node at (0,2.5) {Tokens};
\draw[<-] (input.270) -- ++(0.0,-0.75) node[below] {$\tt{My \ first \ attempt \ at \ writing \ a \ book}$};

\node[block] (llm) at (4.5,1.8) {LLM};
\draw[->] (tokenstext.east) -- (llm.west);

\node[block] (rank) at (11,1.8) {$\begin{array}{c} \hbox{Rank} \\ \hbox{computation} \end{array}$};


\draw[->] (llm.east) -- (rank.west) node[above,pos=0.5] (qtext){$\underline{q}_5 = \Pr(x_5|x_{1}^{4})$};

\draw[->] (llm.east) -- (rank.west) node[below,pos=0.5]
{\begin{tabular}{c|c}
Next word & Probability\\
\hline
reading & 0.3 \\
writing & 0.2 \\
cycling & 0.1 \\
driving & 0.05 \\
$\vdots$ & $\vdots$
\end{tabular}
};

\draw[->] (rank.east) -- ++(1.5,0) node[above,pos=0.5] (ranktext)
{$R_5=1$};


\end{tikzpicture}

%% file: blockdiagram3.tex
\begin{tikzpicture}
[node distance=2cm, block/.style={draw, minimum width=2cm, minimum height=1.5cm}]

\node[block] (input) {Tokenizer};
\draw[->] (input.90) -- ++(0,0.75) node[above] (tokenstext) 
{$\tt{'first',\ 'attempt',\ 'at', \ 'writing'}$};
\node at (0,2.5) {Tokens};
\draw[<-] (input.270) -- ++(0.0,-0.75) node[below] {$\tt{My \ first \ attempt \ at \ writing \ a \ book}$};

\node[block] (llm) at (4.5,1.8) {LLM};
\draw[->] (tokenstext.east) -- (llm.west);

\node[block] (rank) at (11,1.8) {$\begin{array}{c} \hbox{Rank} \\ \hbox{computation} \end{array}$};


\draw[->] (llm.east) -- (rank.west) node[above,pos=0.5] (qtext){$\underline{q}_6 = \Pr(x_6 |x_{2}^{5})$};

\draw[->] (llm.east) -- (rank.west) node[below,pos=0.5]
{\begin{tabular}{c|c}
Next word & Probability\\
\hline
a & 0.7 \\
an & 0.2 \\
the & 0.05 \\
$\vdots$ & $\vdots$
\end{tabular}
};

\draw[->] (rank.east) -- ++(1.5,0) node[above,pos=0.5] (ranktext)
{$R_6=0$};


\end{tikzpicture}

%% file: ranktobits.tex
\begin{tikzpicture}
[node distance=2cm, block/.style={draw, minimum width=2cm, minimum height=1.5cm}]

\node[block] (zip) at (3,0) {Compression (zip)};

\draw[<-](zip.west) -- ++(-2,0) node[above]{
$\begin{array}{c}
\hbox{Sequence of ranks} \\
r_1, r_2, \ldots, r_{N_T}
\end{array}$};

\draw[->] (zip.east) -- ++(1,0) node[above] {$N_b$ bits};

\end{tikzpicture}

%% file: blockdiagram1.tex
\begin{tikzpicture}
[node distance=2cm, block/.style={draw, minimum width=2cm, minimum height=1.5cm}]

\node[block] (input) {Tokenizer};
\draw[->] (input.90) -- ++(0,0.75) node[above] (tokenstext) {Tokens $\ldots,x_{i-M}, \ldots, x_{i-1}$};
\draw[<-] (input.270) -- ++(0.0,-0.75) node[below] {Input sentence $\underline{s}$ with $N_c$ characters};

\node[block] (llm) at (4,1.8) {LLM};
\draw[->] (tokenstext.east) -- (llm.west);

\node[block] (rank) at (11,1.8) {$\begin{array}{c} \hbox{Rank} \\ \hbox{computation} \end{array}$};

\node[block] (zip) at (16,1.8) 
{\begin{tabular}{c}
Losseless \\
Compression
\end{tabular}
};

\draw[->] (llm.east) -- (rank.west) node[above,pos=0.5] (qtext){$\underline{q}_i = \Pr(X_i = j |x_{i-1}^{i-M})$};

\draw[->] (rank.east) -- (zip.west) node[above,pos=0.5] (ranktext)
{$r_i$};

\draw[->] (zip.east) -- ++(1,0) node[above] {$N_b$ bits};

\draw[dashed] (7.5,1.8) -- (7.5,0.5);
\draw[dashed] (7.5,0.5) -- (13.5,0.5);
\draw[dashed] (13.5,0.5) -- (13.5,1.5);
\draw[->,dashed] (13.5,1.5) -- (14.75,1.5);

\end{tikzpicture}

%% file: main.bbl
\begin{thebibliography}{10}

\bibitem{cover1978convergent}
Thomas Cover and Roger King,
\newblock ``A convergent gambling estimate of the entropy of english,''
\newblock {\em IEEE Transactions on Information Theory}, vol. 24, no. 4, pp.
  413--421, 1978.

\bibitem{lutati2023focus}
Shahar Lutati, Itamar Zimerman, and Lior Wolf,
\newblock ``Focus your attention (with adaptive {IIR} filters),'' 2023.

\bibitem{shannon1951prediction}
Claude~E Shannon,
\newblock ``Prediction and entropy of printed english,''
\newblock {\em Bell system technical journal}, vol. 30, no. 1, pp. 50--64,
  1951.

\bibitem{cleary1984data}
John Cleary and Ian Witten,
\newblock ``Data compression using adaptive coding and partial string
  matching,''
\newblock {\em IEEE transactions on Communications}, vol. 32, no. 4, pp.
  396--402, 1984.

\bibitem{goyal2018deepzip}
Mohit Goyal, Kedar Tatwawadi, Shubham Chandak, and Idoia Ochoa,
\newblock ``Deepzip: Lossless data compression using recurrent neural
  networks,''
\newblock {\em arXiv preprint arXiv:1811.08162}, 2018.

\bibitem{touvron2023llama}
Hugo Touvron, Thibaut Lavril, Gautier Izacard, Xavier Martinet, Marie-Anne
  Lachaux, Timothée Lacroix, Baptiste Rozière, Naman Goyal, Eric Hambro,
  Faisal Azhar, Aurelien Rodriguez, Armand Joulin, Edouard Grave, and Guillaume
  Lample,
\newblock ``Llama: Open and efficient foundation language models,'' 2023.

\bibitem{gutenberg_texas}
J.~Frank Dobie,
\newblock {\em Legends of Texas},
\newblock United States, Texas Folk-Lore Society, 1924; Project Gutenberg, May
  25, 2023, 2023,
\newblock \url{https://www.gutenberg.org/ebooks/70859}.

\bibitem{cover1999information}
Thomas~M Cover and Joy~A Thomas,
\newblock {\em Elements of Information Theory},
\newblock Wiley, New York, 1999.

\bibitem{bell1989modeling}
Timothy Bell, Ian~H Witten, and John~G Cleary,
\newblock ``Modeling for text compression,''
\newblock {\em ACM Computing Surveys (CSUR)}, vol. 21, no. 4, pp. 557--591,
  1989.

\bibitem{mackay2003information}
David~JC MacKay,
\newblock {\em Information theory, inference and learning algorithms},
\newblock Cambridge university press, 2003.

\bibitem{kudo2018sentence}
Taku Kudo and John Richardson,
\newblock ``Sentencepiece: {A} simple and language independent subword
  tokenizer and detokenizer for neural text processing,''
\newblock {\em CoRR}, vol. abs/1808.06226, 2018.

\bibitem{text8results}
``text8 results,'' http://mattmahoney.net/dc/textdata.html.

\end{thebibliography}
